\documentclass{appolb}
\usepackage{epsfig}
\usepackage{graphicx}% Include figure files
\usepackage{bm}% bold math
\usepackage{amsmath,epsfig}
\usepackage{natbib}
\usepackage{dcolumn}
\usepackage{graphicx}
\usepackage{amsmath}
\usepackage[hang,small,bf,centerlast]{caption} 
\begin{document}
\title{Sector identification in a set of stock return time series traded at the London Stock Exchange
\thanks{Presented at ...}%
}
\author{C. Coronnello$^{\times}$, M. Tumminello$^{+}$ , F. Lillo$^{\times,+,\dag}$, S.
Miccich\`e$^{\times,+}$ and R.N. Mantegna$^{\times,+}$
\address{$^{\times}$ INFM-CNR, Unit\`a di Palermo, Palermo, Italy}
\address{$^+$ Dipartimento di Fisica e Tecnologie Relative, Universit\`a degli Studi di Palermo, Viale delle Scienze, Edificio 18, I-90128, Palermo, Italy}
\address{$^{\dag}$ Santa Fe Institute, 1399 Hyde Park Road, Santa Fe, NM 87501, USA}
}
\maketitle
\begin{abstract}
We compare some methods recently used in the literature to detect the existence of a certain degree of common behavior of stock returns belonging to the same economic sector. Specifically, we discuss methods based on random matrix theory and hierarchical clustering techniques. We apply these methods to a portfolio of stocks traded at the London Stock Exchange. The investigated time series are recorded both at a daily time horizon and at a 5-minute time horizon. The correlation coefficient matrix is very different at different time horizons confirming that more structured correlation coefficient matrices are observed for long time horizons. All the considered methods are able to detect economic information and the presence of clusters characterized by the economic sector of stocks. However different methods present a different degree of sensitivity with respect to different sectors. Our comparative analysis suggests that the application of just a single method could not be able to extract all the economic information present in the correlation coefficient matrix of a stock portfolio.
\end{abstract}
  
%%%%%%%%%%%%%%%%%%%%%%%%%%%%%%%%%%%%%%%%%%%%%%%%%%%%%%%%%%%%%%%%%%%%%%%%%%%%%%%%%%%%%%%%%%%%%%%%%%%%%%%%%%
%%%%%%%%%%%%%%%%%%%%%%%%%%%%%%%%%%%%%%%%%%%%%%%%%%%%%%%%%%%%%%%%%%%%%%%%%%%%%%%%%%%%%%%%%%%%%%%%%%%%%%%%%%
\section{Introduction}

Multivariate time series are detected and recorded both in experiments and in
the monitoring of a wide number of physical, biological and 
economic systems. A first instrument in the investigation of a multivariate
time series is the correlation matrix.
The study of the properties of the correlation matrix has a direct 
relevance in the investigation of mesoscopic physical systems \cite{Forrester94},
high energy physics \cite{Demasure2003}, information theory and
communication \cite{Moustakas2000,Tworzydlo2002,Skipetrov2003}, investigation
of microarray data in biological systems \cite{maritan,brown,banavar} and
econophysics \cite{Laloux1999,Plerou1999,Mantegna1999,maslov,pafka,sornette,burda}.

Multivariate stock return time series are characterized by a correlation matrix which is carrying information about the economic sectors of the considered stocks \cite{Mantegna1999,Kullmann2000,Bonanno2000,Gopikrishnan2001,Bonanno2001,Kullmann2002,Onnela2002,Giada2002,Bonanno2003,Bonanno2004,Tumminello2005}.

Recent empirical and theoretical analysis have shown that this information can be detected by using a variety of methods. In this paper we review some of these methods based on Random Matrix Theory (RMT) \cite{Gopikrishnan2001}, correlation based clustering \cite{Mantegna1999},  and topological properties of correlation based graphs \cite{Tumminello2005}. The common and different aspects of these methods are discussed by considering the results of an analysis investigating the set of $n=92$ stocks belonging to ``SET 1'' of the London Stock Exchange (LSE). The time period of the time series is the entire 2002 year and the analysis is performed at two different time horizons. Specifically, we investigate the 5-minute time horizon and the daily time horizon to show the differences detected in the structure of the correlation matrix of high frequency and daily returns.

The paper is organized as follows: in Section 2 we discuss the methods used to extract economic information from a correlation matrix of a stock portfolio by using concepts and tools of RMT and hierarchical clustering. The investigated correlation based clustering procedures are the single linkage and average linkage. We also consider a graph obtained by imposing the topological constraint of planarity during its construction along a well defined algorithmic procedure. This graph has been named by authors as the Planar Maximally Filtered Graph (PMFG). In Section 3 we present the empirical results obtained for daily returns of the $92$ stocks belonging to ``SET 1'' of the LSE recorded in 2002. Section 4 presents the the empirical results obtained for 5-minute returns of the same set of data. In Section 5 we draw our conclusions.

%%%%%%%%%%%%%%%%%%%%%%%%%%%%%%%%%%%%%%%%%%%%%%%%%%%%%%%%%%%%%%%%%%%%%%%%%%%%%%%%%%%%%%%%%%%%%%%%%%%%%%%%%%
%%%%%%%%%%%%%%%%%%%%%%%%%%%%%%%%%%%%%%%%%%%%%%%%%%%%%%%%%%%%%%%%%%%%%%%%%%%%%%%%%%%%%%%%%%%%%%%%%%%%%%%%%%
\section{Methods}

In this section we review several methods used to select
part of the content of the correlation coefficient matrix 
which is robust with respect to statistical uncertainty 
and carrying economic information. 

The correlation coefficient between the time evolution of two stock return time series is defined as
\begin{eqnarray}
                \rho_{ij}(\Delta t)=\frac{\langle r_i r_j\rangle -\langle r_i\rangle\langle r_j\rangle}
{\sqrt{(\langle r_i^2\rangle-\langle r_i\rangle^2)(\langle r_j^2\rangle -\langle r_j\rangle^2)}} 
                \qquad \qquad i,j=1, \dots, n
\end{eqnarray} 
where $n$ is the number of stocks, $i$ and $j$ label the stocks, $r_i$ is the logarithmic return defined by $r_i = \ln P_i (t) - \ln P_i (t - \Delta t)$, $P_i (t)$ is the value of the stock price $i$ at the trading time $t$ and $\Delta t$ is the time horizon at which one computes the returns. In this work the correlation coefficient is computed between synchronous return time series. The  correlation coefficient matrix is an $n \times n$ matrix whose elements are the  correlation coefficients $\rho_{ij} (\Delta t)$.

We start our review of methods by discussing the 
application of concepts of RMT which have been used
to select the eigenvalues and eigenvectors of the correlation
matrix less affected by statistical uncertainty. Then
we consider two different correlation based clustering 
procedures. Correlation based clustering procedures are used to 
obtain a reduced number of similarity measures representative of the whole original correlation matrix. The 
filtering procedure associated with a reduction of the
considered similarity measures is typically going from $n(n-1)/2$ distinct elements to a number of similarity measures of the order of $n$.  
The first clustering procedure we consider here is the single linkage clustering method that has been repeatedly
used to detect a hierarchical organization of stocks and the
associated Minimum Spanning Tree (MST) and PMFG.
The PMFG is a recently introduced 
graph extending the number of similarity measures
associated to the graph with respect to the ones present in the
MST. This extension of considered links is done by conserving
the same hierarchical tree of the MST \cite{Tumminello2005}.
The second clustering procedure is the average 
linkage which provides a different taxonomy and the last one is
the PMFG.

%%%%%%%%%%%%%%%%%%%%%%%%%%%%%%%%%%%%%%%%%%%%%%%%%%%%%%%%%%%%%%%%%%%%%%%%%%%%%%%%%%%%%%%%%%%%%%%%%%%%%%%%%%
\subsection{Random Matrix Theory} \label{RMT}

Random Matrix Theory \cite{Metha90} was originally developed in nuclear physics and then applied to many different fields. In the context of asset portfolio management RMT is useful because it allows to compute the effect of statistical uncertainty in the estimation of the correlation matrix. 
Suppose that the $n$ assets are described by $n$ time series of $T$ time records and that the returns are  independent Gaussian random variables with zero mean and variance $\sigma^2$. The correlation matrix of this set of variables in the limit $T\to \infty$ is simply the identity matrix. When $T$ is finite the correlation matrix will in general be different from the identity matrix. RMT allows to prove that in the limit $T,n \to \infty$, with a fixed ratio $Q=T/n \geq 1$, the eigenvalue spectral density of the covariance matrix is given by
\begin{equation}
\rho(\lambda)=\frac{Q}{2\pi\sigma^2\lambda}\sqrt{(\lambda_{max}-\lambda)
(\lambda-\lambda_{min})}, 
\label{zerofactor}
\end{equation}
where $\lambda_{min}^{max}=\sigma^2 (1+1/Q\pm 2\sqrt{1/Q})$. 
The spectral density is different from zero in the interval $]\lambda_{min},\lambda_{max}[$. In the case of a correlation matrix it is $\sigma^2=1$.
The spectrum described by Eq.~\ref{zerofactor} is different from $\delta(\lambda-1)$ which is expected by an identity correlation matrix. In other words RMT quantifies the role of the finiteness of the length of the time series on the spectral properties of the correlation matrix. 

RMT has been applied to the investigation of correlation matrices of financial asset returns \cite{Laloux1999,Plerou1999} and it has been shown that the spectrum of a typical portfolio can be divided in three classes of eigenvalues. The largest eigenvalue is totally incompatible with Eq.~\ref{zerofactor} and describes the common behavior of the stocks composing the portfolio. This fact leads to another working hypothesis that the part of correlation matrix which is orthogonal to the eigenvector corresponding to the first eigenvalue is random. This amounts to quantify the variance of the part not explained by the highest eigenvalue as $\sigma^2=1-\lambda_1/n$ and to use this value in Eq.~\ref{zerofactor} to compute $\tilde\lambda_{min}$ and $\tilde\lambda_{max}$. Under this assumption, previous studies have shown that a fraction of the order of few percent of the eigenvalues are also incompatible with the RMT because they fall outside the interval   $]\tilde\lambda_{min},\tilde\lambda_{max}[$ computed with the value of $\sigma$ taking into account the behavior of the first eigenvalue. These eigenvalues probably describe economic information stored in the correlation matrix. The remaining large part of the eigenvalues is between  $\tilde\lambda_{min}$ and $\tilde\lambda_{max}$ and thus one cannot say whether any information is contained in the corresponding eigenspace. 

The fact that by using RMT it is possible, under certain assumptions, to identify the  part of the correlation matrix containing economic information suggested some authors to use RMT for showing that some selected eigenvectors, i.e. eigenvectors associated to eigenvalues not explained by RMT, describe economic sectors. Specifically the suggested method \cite{Gopikrishnan2001} is the following. One computes the correlation matrix and finds the spectrum ranking the eigenvalues such that $\lambda_k>\lambda_{k+1}$. The  eigenvector corresponding to $\lambda_k$ is denoted ${\bf u}^k$. The set of investigated stocks is partitioned in $S$ sectors $s=1,2,...,S$ according to their economic activity (for example by using classification codes such as the one of the Standard Industrial Classification code or Forbes). One then defines a $S\times n$ projection matrix ${\bf P}$ with elements $P_{si}=1/n_s$ if stock $i$ belongs to sector $s$ and $P_{si}=0$ otherwise. Here $n_s$ is the number of stocks belonging to sector $s$. For each eigenvector ${\bf u}^k$ one computes
\begin{equation}
X^k_s\equiv\sum_{i=1}^n P_{si}[u_i^k]^2
\label{projectioneq}
\end{equation}
This number gives a measure of the role of a given sector $s$ in explaining the composition of eigenvector ${\bf u}^k$. Thus when a given eigenvector has a large value of $X^k_s$ for only one (or few) sector $s$, one can conclude that the eigenvector describes that economic sector. Note that this method requires the {\it a priori} knowledge of the sector for each stock in order to be implemented.

%%%%%%%%%%%%%%%%%%%%%%%%%%%%%%%%%%%%%%%%%%%%%%%%%%%%%%%%%%%%%%%%%%%%%%%%%%%%%%%%%%%%%%%%%%%%%%%%%%%%%%%%%%
\subsection{Hierarchical Clustering Methods}

Another approach used to detect the information associated to the correlation matrix is given by the correlation based
hierarchical clustering analysis. Consider a set of $n$ objects and suppose that a similarity measure, e.g. the correlation coefficient, between pairs of elements is defined. Similarity measures can be written in a $n\times n$ similarity matrix. The hierarchical clustering methods allow to hierarchically organize the elements in clusters. The result of the procedure is a rooted tree or dendrogram giving a quantitative description of the clusters thus obtained. It is worth noting that hierarchical clustering methods can as well be applied to distance matrices.

A large number of hierarchical clustering procedures can be found in the literature. For a review about the classical techniques see for instance Ref. \cite{Anderberg}. In this paper we focus out attention on the Single Linkage Cluster Analysis (SLCA), which was introduced in finance in Ref. \cite{Mantegna1999} and the Average Linkage Cluster Analysis (ALCA).
    
%%%%%%%%%%%%%%%%%%%%%%%%%%%%%%%%%%%%%%%%%%%%%%%%%%%%%%%%%%%%%%%%%%%%%%%%%%%%%%%%%%%%%%%%%%%%%%%%%%%%%%%%%%
\subsubsection{Single Linkage Correlation Based Clustering} \label{SLCA}

The Single Linkage Cluster Analysis is a filtering procedure based on the estimation of the subdominant ultrametric distance \cite{Rammal1986} associated with a metric distance obtained from the correlation coefficient matrix of a set of $n$ stocks. This procedure, already used in other fields, allows to extract a
MST and a hierarchical tree from a correlation coefficient
matrix by means of a well defined algorithm known as nearest neighbor single linkage
clustering algorithm \cite{Mardia}. This methodology allows to reveal both topological (through the MST)
and taxonomic (through the hierarchical tree) aspects of the correlation present among stocks.

The MST is obtained by selecting a relevant part of the information which is
present in the correlation coefficient matrix of the time series of stock returns. 
In the present study this is done (i) by determining the synchronous correlation coefficient of the
difference of logarithm of stock price computed at a selected time horizon, (ii) by
calculating a metric distance between all the pair of stocks and (iii) by selecting
the subdominant ultrametric distance associated to the considered metric distance.
The subdominant ultrametric is the ultrametric structure closest to the original
metric structure \cite{Rammal1986}.

A metric distance between pair of stocks can be rigorously determined \cite{Gower1966} by defining
\begin{eqnarray}
                d_{ij}=\sqrt{2 (1-\rho_{ij})} \label{distance}
\end{eqnarray} 
With this choice $d_{ij}$ fulfills the three axioms of a metric ­ (i) $d_{ij} = 0$ if
and only if $i = j$ ; (ii) $d_{ij} = d_{ji}$ and (iii) $d_{ij} \le d_{ik} + d_{kj}$. The distance matrix ${\bf{D}}$ is then used to determine the MST connecting the $n$ stocks.

The MST is a graph without loops connecting all the $n$ nodes with the shortest $n - 1$ links amongst all the links between the nodes. The selection of these $n-1$ links is done according to some widespread algorithm \cite{Papadimitriou82} and can be summarized as follows:
\begin{enumerate} 
\item Construct an ordered list of pair of stocks $L_{ord}$, by ranking all the possible pairs according to their distance $d_{ij}$. The first pair of $L_{ord}$ has the shortest distance.
\item The first pair of $L_{ord}$ gives the first two elements of the MST and the link between them. 
\item The construction of the MST continues by analyzing the list $L_{ord}$. At each successive stage, a pair of elements is selected from $L_{ord}$ and the corresponding link is added to the MST only if no loops are generated in the graph after the link insertion.
\end{enumerate}
Different elements of the list are therefore iteratively included in the MST starting from the first two elements of $L_{ord}$. As a result, one obtains a graph with $n$ vertices and $n-1$ links. For a didactic description of the method used to obtain the MST one can consult Ref. \cite{MS00}

In Ref. \cite{Gower1969} the procedure briefly sketched above has been shown to provide a MST which is associated to the same hierarchical tree of the SLCA. In this procedure, at each step, when two elements or one element and a cluster or two clusters $p$ and $q$ merge in a wider single cluster $t$, the distance $d_{tr}$ between the new cluster $t$ and any cluster $r$ is recursively determined as follows:  
\begin{equation}
                d_{tr}=\min \{ d_{pr},d_{qr}\}
\end{equation}
thus indicating that the distance between any element of cluster $t$ and any element of cluster $r$ is the shortest distance between any two entities in clusters $t$ and $r$. By applying iteratively this procedure $n-1$ of the $n(n-1)/2$ distinct elements of the original correlation coefficient matrix are selected. 

The distance matrix obtained by applying the SLCA is an ultrametric matrix comprising $n-1$ distinct selected elements. The ultrametric distance $d_{ij}^<$ between element $i$ belonging to cluster $t$ and element $j$ belonging to cluster $r$ is therefore defined as the distance between clusters $t$ and $r$. Ultrametric distances $d_{ij}^<$ are distances satisfying the inequality $d^<_{ij} \le  \max \{d^<_{jk},d^<_{kj}\}$ stronger than the customary triangular inequality $d_{ij} \le  d_{ik} + d_{kj}$ \cite{Rammal1986}. The SLCA has associated an ultrametric correlation matrix which is the subdominant ultrametric matrix of the original correlation coefficient matrix. It can be obtained starting from the ultrametric distances $d_{ij}^<$ and making use of Eq. \ref{distance}.

The MST allows to obtain, in a direct and essentially unique way, the subdominant ultrametric distance matrix ${\bf{D}}^< $ and the hierarchical organization of the elements of the investigated data set. In Ref. \cite{TumminelloManuscript} it is proved that the ultrametric correlation matrix obtained by the SLCA is always positive definite when all the elements of the obtained ultrametric correlation matrix are non negative. This condition is rather common in financial data. 

The effectiveness of the SLCA in pointing out the hierarchical structure of the investigated portfolio has been shown by several studies  \cite{Mantegna1999,Bonanno2000,Bonanno2001,Kullmann2002,Onnela2002,Bonanno2003,Bonanno2004,Micciche2003,Dimatteo2004}.

%%%%%%%%%%%%%%%%%%%%%%%%%%%%%%%%%%%%%%%%%%%%%%%%%%%%%%%%%%%%%%%%%%%%%%%%%%%%%%%%%%%%%%%%%%%%%%%%%%%%%%%%%%
\subsubsection{Average Linkage Correlation Based Clustering} \label{ALCA}

The Average Linkage Cluster Analysis  is a hierarchical clustering procedure \cite{Anderberg} that can be described by considering either a similarity or a distance measure. Here we consider the distance matrix ${\bf{D}}$. The following procedure performs the ALCA giving as an output a rooted tree and an ultrametric matrix ${\bf{D}}^<$ of elements $d^<_{ij}$:
\begin{enumerate} 
\item Set ${\bf{T}}$ as the matrix of elements such that ${\bf{T}} = {\bf{D}}$.
\item Select the minimum distance $t_{hk}$ in the distance matrix ${\bf{T}}$. Note that after the first step of construction $h$ and $k$ can be simple elements (i.e. clusters of one element each) or clusters (sets of elements).
\item Merge cluster $h$ and cluster $k$ into a single cluster, say $h$. The merging operation identifies a node in the rooted tree connecting clusters $h$ and $k$ at the distance $t_{hk}$. Furthermore to obtain the ultrametric matrix it is sufficient that $\forall \, i \in h$ and $\forall \, j \in k$ one sets $d^<_{ij}=d^<_{ji}=t_{hk}$. 
\item Redefine the matrix ${\bf{T}}$:
\begin{eqnarray} \label{negspin}
\left \{  \begin{aligned}
        &   t_{hj}= \frac{N_h\,t_{hj}+N_k \, t_{kj}}
                               {N_h+N_k}  & 
                           ~~~~\text{ if } j\neq h \,{\rm{and}} \, j\neq k\\
        &                 \nonumber \\
        &    t_{ij}=t_{ij} & 
                           ~~~~\text{ otherwise, }\\
\end{aligned} \right.
\end{eqnarray}
where $N_h$ and $N_k$ are the number of elements belonging respectively to the cluster $h$ and to the cluster $k$. Note that if the dimension of ${\bf{T}}$ is $m \times m$ then the dimension of the redefined ${\bf{T}}$ is $(m-1) \times (m-1)$ because of the merging of clusters $h$ and $k$ into the cluster $h$.
\item If the dimension of ${\bf{T}}$ is bigger than one then go to step 2 else Stop.   
\end{enumerate}
By replacing point $4$ of the above algorithm with the following item\\

\indent $4.$  Redefine the matrix ${\bf{T}}$:
\begin{equation}\label{negspin2}
\left \{ \begin{aligned}
        &  t_{hj}= Min \left[t_{hj}, t_{kj}\right] & 
                           ~~~~\text{ if } j\neq h \,{\rm{and}} \, j\neq k \nonumber\\
        &  t_{ij}=t_{ij} & 
                           ~~~~\text{ otherwise, }\\
\end{aligned} \right.
\end{equation}
one obtains an algorithm performing the SLCA which is therefore equivalent to the one described in the previous section. The algorithm can be easily adapted for working with similarities instead of distances. It is just enough to exchange the distance matrix ${\bf{D}}$ with a similarity matrix (for instance the correlation matrix) and replace the search for the minimum distance in the matrix ${\bf{T}}$ in point $2$ of the above algorithm with the search for the maximal similarity. 

It is worth noting that the ALCA can produce different hierarchical trees depending on the use of a similarity matrix or a distance matrix. More precisely, different dendrograms can result for the ALCA due to the non linearity of the transformation of Eq. \ref{distance}. This problem does not arise in the SLCA because Eq. \ref{distance} is a monotonic transformation and therefore it does not affect the search for the minimum (or maximum for the similarity).

%%%%%%%%%%%%%%%%%%%%%%%%%%%%%%%%%%%%%%%%%%%%%%%%%%%%%%%%%%%%%%%%%%%%%%%%%%%%%%%%%%%%%%%%%%%%%%%%%%%%%%%%%%
\subsection{The Planar Maximally Filtered Graph}

The Planar Maximally Filtered Graph has been introduced in a recent paper \cite{Tumminello2005}. The basic idea is to obtain a graph that retains the same hierarchical properties of the MST, i.e. the same hierarchical tree of SLCA, but allowing a greater number of links and more complex topological structures than the MST, such as loops and cliques. Such a graph is obtained by relaxing the topological constraint of the MST construction protocol of section \ref{SLCA} according to which no loops are allowed in a tree. Specifically, in the PMFG a link can be included in the graph if and only if the graph with the new link included is still planar. A graph is planar if and only if it can be drawn on a plane (infinite in principle) without edge crossings \cite{West}. \\

The first difference between MST and PMFG is about the number of links, which is $n-1$ in the MST and $3(n-2)$ in the PMFG. Furthermore loops and cliques are allowed in the PMFG. A clique of $r$ elements, r-cliques, is a subgraph of $r$ elements where each element is linked to each other. Because of the Kuratowski's theorem \cite{West} only 3-cliques and 4-cliques are allowed in the PMFG. The study of 3-cliques and 4-cliques is relevant for understanding the strength of clusters in the system \cite{Tumminello2005} as we will see below in the empirical applications.\\

Concerning the hierarchical structure associated to the PMFG it has been shown in Ref. \cite{Tumminello2005} that at any step of construction of the MST and PMFG, if two elements are connected via at least one path in one of the considered graphs, then they also are connected in the other one. This statement implies that i) the MST is always contained in the PMFG and ii) the hierarchical tree associated to  both the MST and PMFG is the one obtained from the SLCA.\\

In summary the PMFG is a graph retaining more information about the system than the MST, the information being stored in the included new links and in the new topological structures allowed. i.e. loops and cliques.\\

%%%%%%%%%%%%%%%%%%%%%%%%%%%%%%%%%%%%%%%%%%%%%%%%%%%%%%%%%%%%%%%%%%%%%%%%%%%%%%%%%%%%%%%%%%%%%%%%%%%%%%%%%%
%%%%%%%%%%%%%%%%%%%%%%%%%%%%%%%%%%%%%%%%%%%%%%%%%%%%%%%%%%%%%%%%%%%%%%%%%%%%%%%%%%%%%%%%%%%%%%%%%%%%%%%%%%
\section{Empirical Results: Daily Data}
In the present section we apply the selected methods to a set of stocks traded at the LSE. These 
stocks are highly capitalized stocks and they belong to $11$ different economic sectors.

%%%%%%%%%%%%%%%%%%%%%%%%%%%%%%%%%%%%%%%%%%%%%%%%%%%%%%%%%%%%%%%%%%%%%%%%%%%%%%%%%%%%%%%%%%%%%%%%%%%%%%%%%%
\subsection{The Data Set}

We investigate the statistical properties of price returns for $n=92$ highly traded stocks belonging to the SET1 segment of the LSE market {\tt www.londonstockexchange.com}. In particular, we consider electronic transactions occurred in year 2002. The empirical data are taken from the ``Rebuild Order Book'' database, maintained by the LSE.

For each of the 92 stocks considered, the trading activity has been defined in terms of the total number of transactions (electronic and manual) occurred in 2002. Most of the transactions, a mean value of $75 \%$ for the 92 stocks, are of the electronic type.

For each stock and for each trading day we consider the time series of stock price recorded transaction by transaction. Since transactions for different stocks do not happen simultaneously, we divide each trading day (lasting $8^h~30'$) into intervals of 5-minute each. For each trading day, we define $103$ intraday stock price proxies $P_i(t_k)$, with $k=1, \cdots, 103$. The proxy is defined as the transaction price detected nearest to the end of the interval (this is one possible way to deal with high-frequency financial data \cite{Dacorogna2001}). By using these proxies, we perform the price returns $r_i = \ln P_i (t) - \ln P_i (t - \Delta t)$ at time horizons of $\Delta t=5$ minute and $\Delta t$ equal to one trading day. In the case of a daily time horizon the returns are computed as the difference of the logarithms of the closure prices of each successive trading day. In the case of $\Delta t=5$ minute, the returns are always computed as the difference of the logarithms of prices which belong to the same trading day.

To each of the 92 selected stocks an economic sector of activity can be associated according to the classification scheme used in the web--site {\tt{www.euroland.com}}. The relevant economic sectors are reported in Table \ref{classification}, together with the number of stocks belonging to each of them (third column). 

\begin{table}
\begin{center}
\caption{Economic sectors of activity for 92 highly traded stocks belonging to the SET1 segment of the LSE. The classification is done according to the methodology used in the web--site {\tt{www.euroland.com}}. The second column contains the economic sector and the third column contains the number of stocks belonging to the sector.} \label{classification}
\vspace{.5 cm}
\begin{tabular}{||c|l|c||}
\hline
                     & ${\rm{SECTOR}}$      ~&~{\rm{NUMBER}}~\cr \hline
                  1  & Technology            & $ 4$          \\
                  2  & Financial             & $20$          \\
                  3  & Energy                & $ 3$          \\
                  4  & Consumer non-Cyclical & $12$          \\
                  5  & Consumer Cyclical     & $10$          \\
                  6  & Healthcare            & $ 6$          \\
                  7  & Basic Materials       & $ 5$          \\
                  8  & Services              & $19$          \\
                  9  & Utilities             & $ 6$          \\
                 10  & Capital Goods         & $ 5$          \\
                 11  & Transportation        & $ 2$          \cr \hline
\end{tabular}
\end{center}
\end{table} 

%%%%%%%%%%%%%%%%%%%%%%%%%%%%%%%%%%%%%%%%%%%%%%%%%%%%%%%%%%%%%%%%%%%%%%%%%%%%%%%%%%%%%%%%%%%%%%%%%%%%%%%%%%
\subsection{Random Matrix Theory} \label{RMT1day}

For a time horizon of one trading day the largest eigenvalue is $\lambda_1=36.0$ clearly incompatible with RMT and suggesting a driving factor common to all the stocks. This is usually interpreted to be the ``market mode" as described in widespread market models, such as the Capital Asset Pricing Model. The analysis of the components of the corresponding eigenvector confirms this interpretation. In fact the mean component of the first eigenvector is $0.102$ and the standard deviation is $0.022$ showing that all the stocks contribute in a similar way to the eigenvector ${\bf u}^1$.

In our data $Q=T/n=2.71$ and the threshold value $\lambda_{max}$ without taking into account the first eigenvalue is $\lambda_{max}=2.58$. This implies that RMT considers as signal only the first two eigenvalues $\lambda_1$ and $\lambda_2=4.58$. On the other hand if we remove the contribution of the first eigenvalue with the procedure discussed in section \ref{RMT} we get $\tilde\lambda_{max}=1.57$, indicating that the first $6$ eigenvalues could contain economic information. This result shows the importance of taking into account the role of the first eigenvalue.

\begin{figure}[ptb]
\begin{center}
\includegraphics[scale=.3]{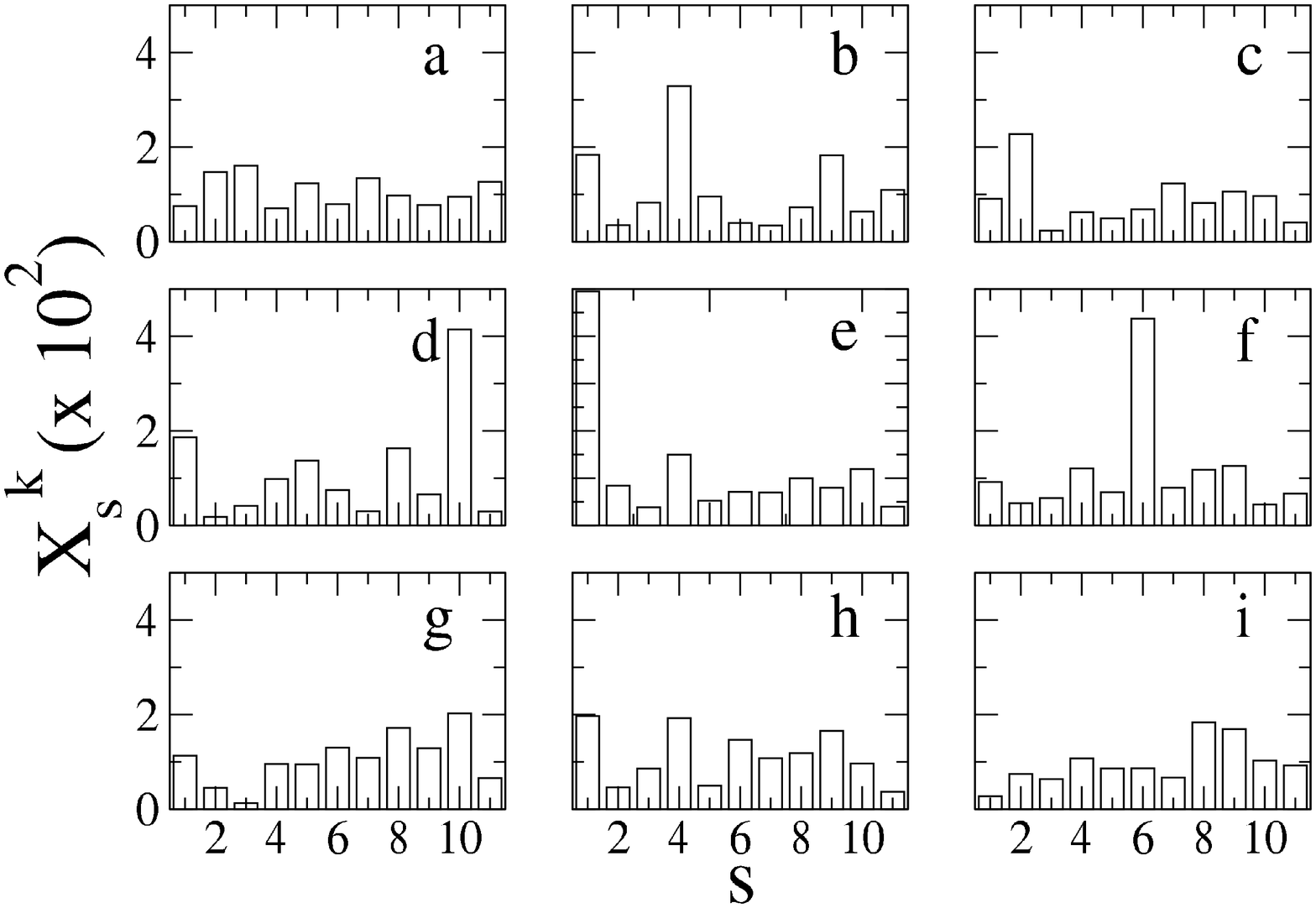}
\caption{
Contribution  $X^k_s$ of Eq.~\ref{projectioneq} for the first (a), second (b), third (c), sixth (f), seventh (g), eighth (h) and ninth (i) eigenvectors of the correlation matrix of daily returns of $92$ LSE stocks. Panel d) shows $X^k_s$ for the linear combination $({\bf u}^4+{\bf u}^5)/\sqrt{2}$ and panel e)  for the linear combination $({\bf u}^4-{\bf u}^5)/\sqrt{2}$. The order of sectors is the same as in Table \ref{classification}.}
\label{projection1day}
\end{center}
\end{figure}

Figure~\ref{projection1day} shows $X^k_s$ of Eq.~\ref{projectioneq} of the first $9$ eigenvalues. Panel (a) shows that all the sectors contribute roughly in a similar way to the first eigenvector. On the other hand eigenvectors $2$, $3$, and $6$ are characterized by one prominent sector. Specifically, the second eigenvector shows a large contribution from the sector Consumer non-Cyclical ($s=4$), the third eigenvector has a significant contribution from the Financial sector ($s=2$), and the sixth eigenvector shows a prominent peak for the stocks of the sector Healthcare ($s=6$). The fourth ($\lambda_4=1.79$) and fifth ($\lambda_5=1.72$) eigenvalues are very close and a plot of $X^k_s$ for the corresponding eigenvectors shows two peaks corresponding to the sectors Capital Goods and Technology. By following a line of reasoning presented in Ref. \cite{Gopikrishnan2001} a possible explanation is that the noise due to the measurement favors the mixing of these two groups. In  support of this hypothesis in panel d) we show $X^k_s$ for the linear combination $({\bf u}^4+{\bf u}^5)/\sqrt{2}$ and in panel e)  for the linear combination $({\bf u}^4-{\bf u}^5)/\sqrt{2}$. Panel d) has a large peak for the sector Capital Goods ($s=10$) and in panel e) the peak is associated to the Technology sector ($s=1$). Finally the seventh, eighth and ninth eigenvector do not  show significant peaks, indicating that probably these are eigenvectors of eigenvalues strongly affected by statistical uncertainty (``noise dressed''). It is interesting to note that the RMT after subtracting the contribution of the largest eigenvector as described above predicts that the first $6$ eigenvalues are deviating, i.e. are outside the noise region. This is the same number one obtains from the analysis of eigenvector component.

%%%%%%%%%%%%%%%%%%%%%%%%%%%%%%%%%%%%%%%%%%%%%%%%%%%%%%%%%%%%%%%%%%%%%%%%%%%%%%%%%%%%%%%%%%%%%%%%%%%%%%%%%%
\subsection{Single Linkage Correlation Based Clustering} \label{SLCA1day}

\begin{figure}[ptb]
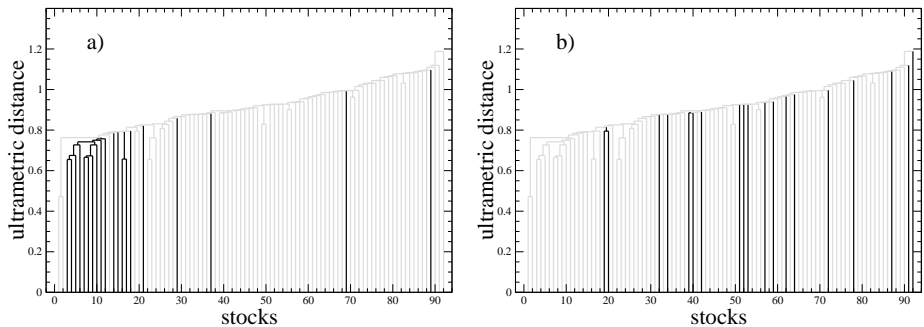

\begin{center}
              \hbox{
              \includegraphics[scale=0.25]{HT_sl_day_financial.eps}
              \hspace{.05 cm}
              \includegraphics[scale=0.25]{HT_sl_day_services.eps}
              }
\end{center}
              \caption{Hierarchical tree obtained by using the SLCA starting from the daily price returns of 92 highly traded stocks belonging to the SET1 segment of the LSE. Only electronic transactions occurred in year 2002 are considered. In panel a) the Financial economic sector is highlighted. In panel b) the Services economic sector is highlighted.} \label{HT_sl_day}
\end{figure}

\begin{figure}[ptb]
\begin{center}
              \includegraphics[scale=0.35]{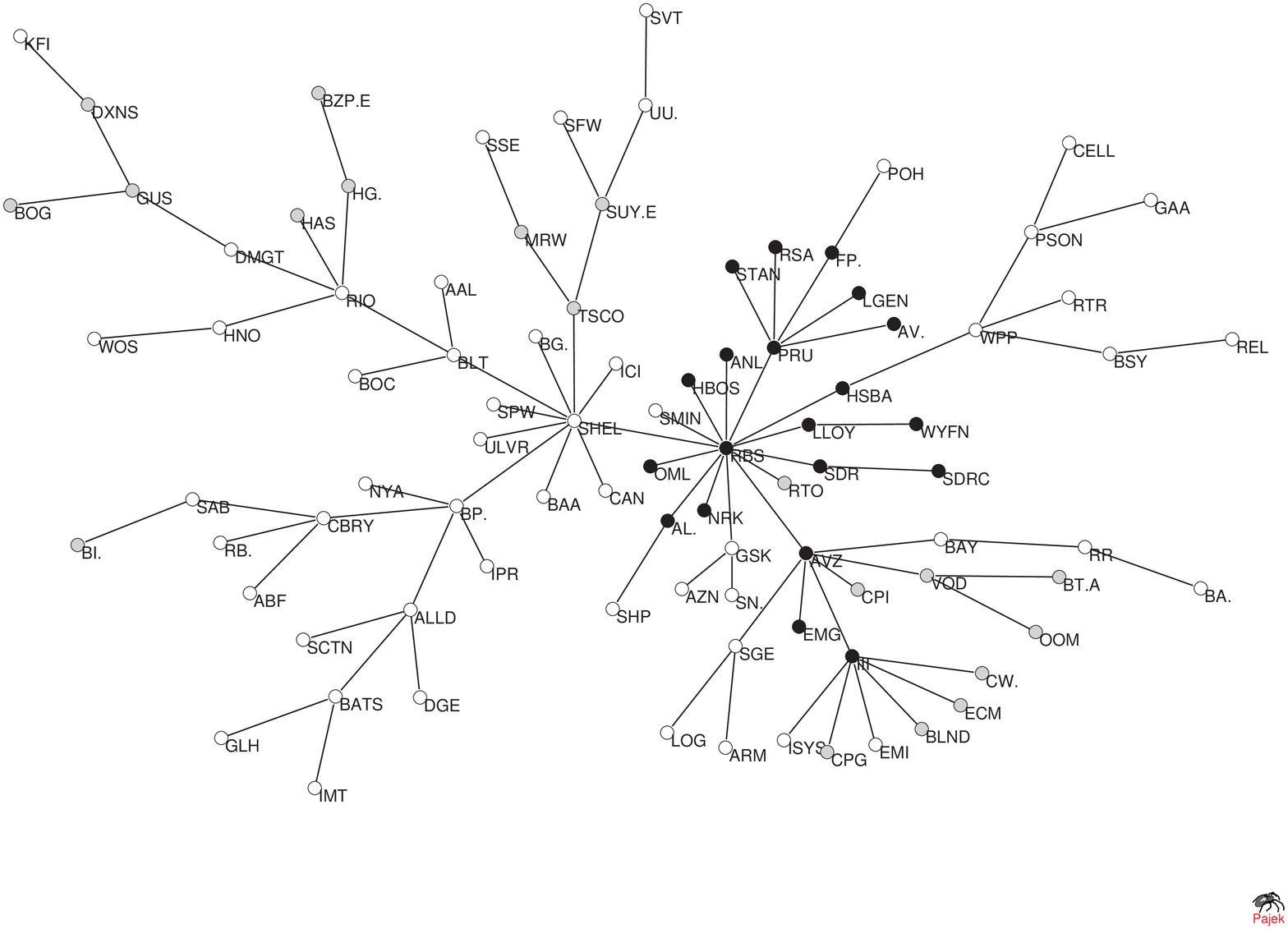}
              \caption{MST obtained starting from the daily price returns of 92 highly traded stocks belonging to the SET1 segment of the LSE. Only electronic transactions occurred in year 2002 are considered. The Financial economic sector (black) and the Services (gray) economic sector are highlighted. It is evident the existence of two stocks that behave as hubs. One of them is RBS, which gathers 14 stocks, 10 of which belong to the Financial sector. The other hub is SHEL which gathers 10 stocks.} \label{MST_sl_day}
\end{center}
\end{figure}

The results obtained by using the SLCA for the daily returns are summarized in Fig. \ref{HT_sl_day} and Fig. \ref{MST_sl_day} that show the hierarchical tree and the MST, respectively.

The hierarchical tree shows that there exists a significant level of correlation in the market, and in some case clustering can be observed. In particular, the first two stocks on the left of Fig. \ref{HT_sl_day}, {\em{Shell}} (SHEL) and {\em{British Petroleum}} (BP), belonging to the Energy sector, are linked together at an ultrametric distance $d^<=0.47$ corresponding to a correlation coefficient as high as $\rho=0.89$. However, the third stock belonging to the Energy economic sector (stock 13), which is {\em{British Gas}} (BG), is not linked to the other two but it is linked to stocks belonging to the Financial sector. We focus here our attention on the two sectors with the largest number of stocks, which are the Financial sector ($s=2$) and the Services sector ($s=8$). Panel a) of Fig. \ref{HT_sl_day} gives an example in which some of the stocks belonging to the same economic sector, e.g. Financial, are clustered together. In fact, a cluster including 10 stocks from position 3 to position 12 can be observed. Panel b) of Fig. \ref{HT_sl_day} gives an example of the opposite case in which stocks belonging to the same economic sector, e.g. Services, are poorly clustered. In fact, only two small clusters of two stocks are formed.
%In appendix \ref{SL} we give the list of all stocks ordered according to hierarchical tree, i.e. according to the $x$-axis of Fig. \ref{HT_sl_day}.

The MST confirms the above results. In Fig. \ref{MST_sl_day} the stocks belonging to the Financial economic sector (black circles) and the Services (gray circles) are indicated. An inspection of Fig. \ref{MST_sl_day} shows that the stocks of the Financial sector cluster around {\em{Royal Bank of Scotland}} (RBS) whereas stocks of the Services sector are present in different branches of the MST. The MST also gives an additional information about the topology of the network. In fact, it is evident from the figure that there are two stocks that behave as hub. One of them is RBS, which gathers 14 stocks, 10 of which belong to the Financial sector. The other hub is SHEL which gathers 10 stocks, among which we find BG and BP. 

%%%%%%%%%%%%%%%%%%%%%%%%%%%%%%%%%%%%%%%%%%%%%%%%%%%%%%%%%%%%%%%%%%%%%%%%%%%%%%%%%%%%%%%%%%%%%%%%%%%%%%%%%%
\subsection{Average Linkage Correlation Based Clustering} \label{ALCA1day}

\begin{figure}
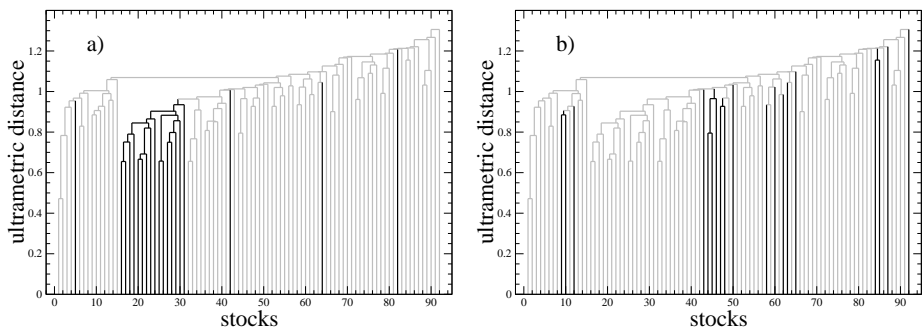
 
\begin{center} 
              \hbox{ 
              \includegraphics[scale=0.25]{HTDistAL_LSE92_2002_day_financial} 
              \hspace{.05 cm} 
              \includegraphics[scale=0.25]{HTDistAL_LSE92_2002_day_services} 
              } 
              \caption{Dendrogram associated to the ALCA performed on daily returns of a portfolio of 92 stocks traded in the LSE in 2002. Panel a): The black lines are identifying stocks belonging to the Financial Sector. Panel b): The black lines are identifying stocks belonging to the Services Sector} 
\label{daydendroaverage} 
\end{center} 
\end{figure}

In this subsection we analyze the dendrogram of Fig. \ref{daydendroaverage} obtained by applying the ALCA to the correlation based distance matrix of the daily returns. Once again, to provide representative examples we focus our attention to the two sectors with the largest number of stocks. As in Fig. \ref{HT_sl_day} in panel a) of Fig. \ref{daydendroaverage} the black lines are identifying stocks of the Financial sector. It can be seen from the figure that most of the stocks (specifically, 16 out of 20) belonging to the Financial sector cluster together at a low level distance ($d \sim 0.85$). Exceptions (referring to black lines outside the cluster in panel a) from the left to the right) are {\em{Northern Rock}} (NRK), {\em{Royal \& Sun Alliance}} (RSA), {\em{Canary Wharf Group}} (WYFN) and {\em{Man Group}} (EMG). Interestingly, RSA, WYFN and EMG are distant from the observed cluster also when considering the SLCA, as shown in panel a) of Fig. \ref{HT_sl_day} at position 37, 69 and 89, respectively. In panel b) of Fig. \ref{ALCA1day} the black lines are identifying the $19$ stocks belonging to the Services sector. In this case just an intra-sector cluster of 3 stocks is detected, specifically the one composed by {\em{Vodafone Group}} (VOD), {\em{mmO$_2$}} (OOM) and {\em{British Telecom}} (BT-A), the corresponding stock numbers in panel b) being respectively 44, 45 and 46. 

\begin{figure}[ptb]
\begin{center}
              \hbox{
              \includegraphics[scale=0.25]{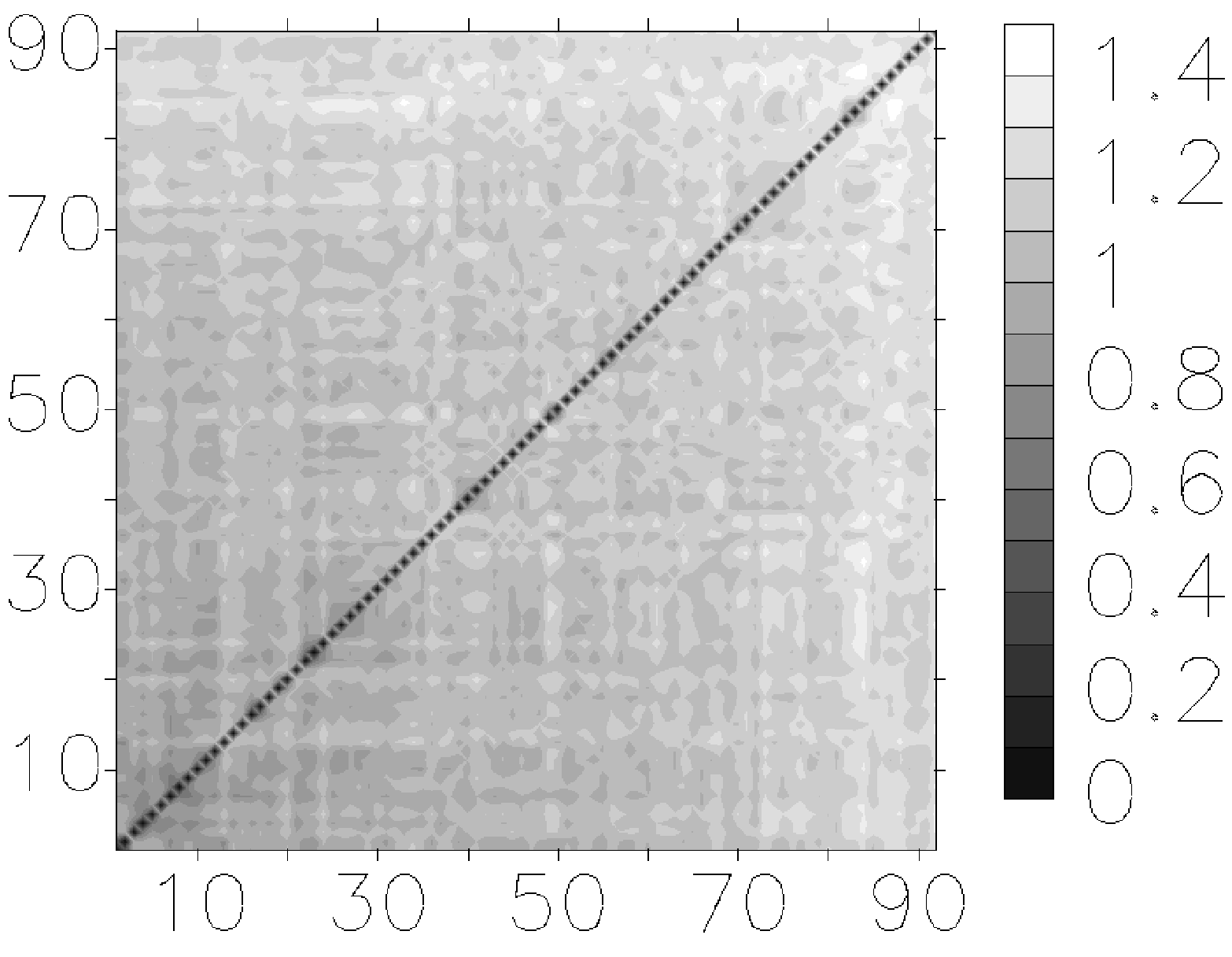}
              \hspace{.05 cm}
              \includegraphics[scale=0.25]{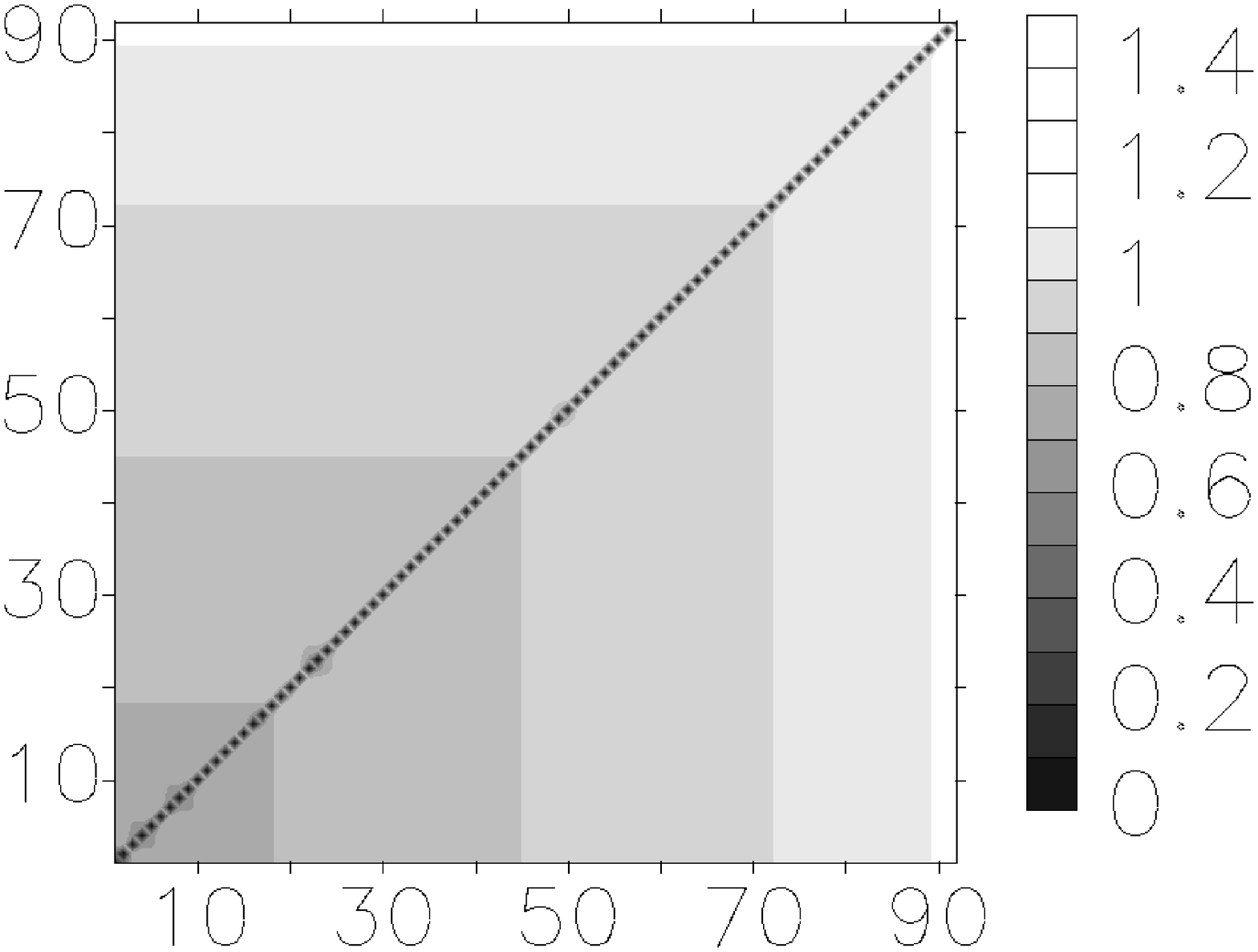}
              }
              \caption{Contour plots of the original correlation matrix (panel a)) and of the one associated to the ultrametric distance (panel b)) obtained by using the SLCA for the daily price returns of 92 highly traded stocks belonging to the SET1 segment of the LSE. Only electronic transactions occurred in year 2002 are considered. Here the stocks are identified by a numerical label ranging from 1 to 92 and ordered according to the hierarchical tree of Fig. \ref{HT_sl_day}. The figure gives a pictorial representation of the amount of information which is filtered out by using the SLCA.} \label{MATRIX_sl_day}
\end{center}
\end{figure}

\begin{figure}[ptb]
\begin{center}
              \hbox{
              \includegraphics[scale=0.25]{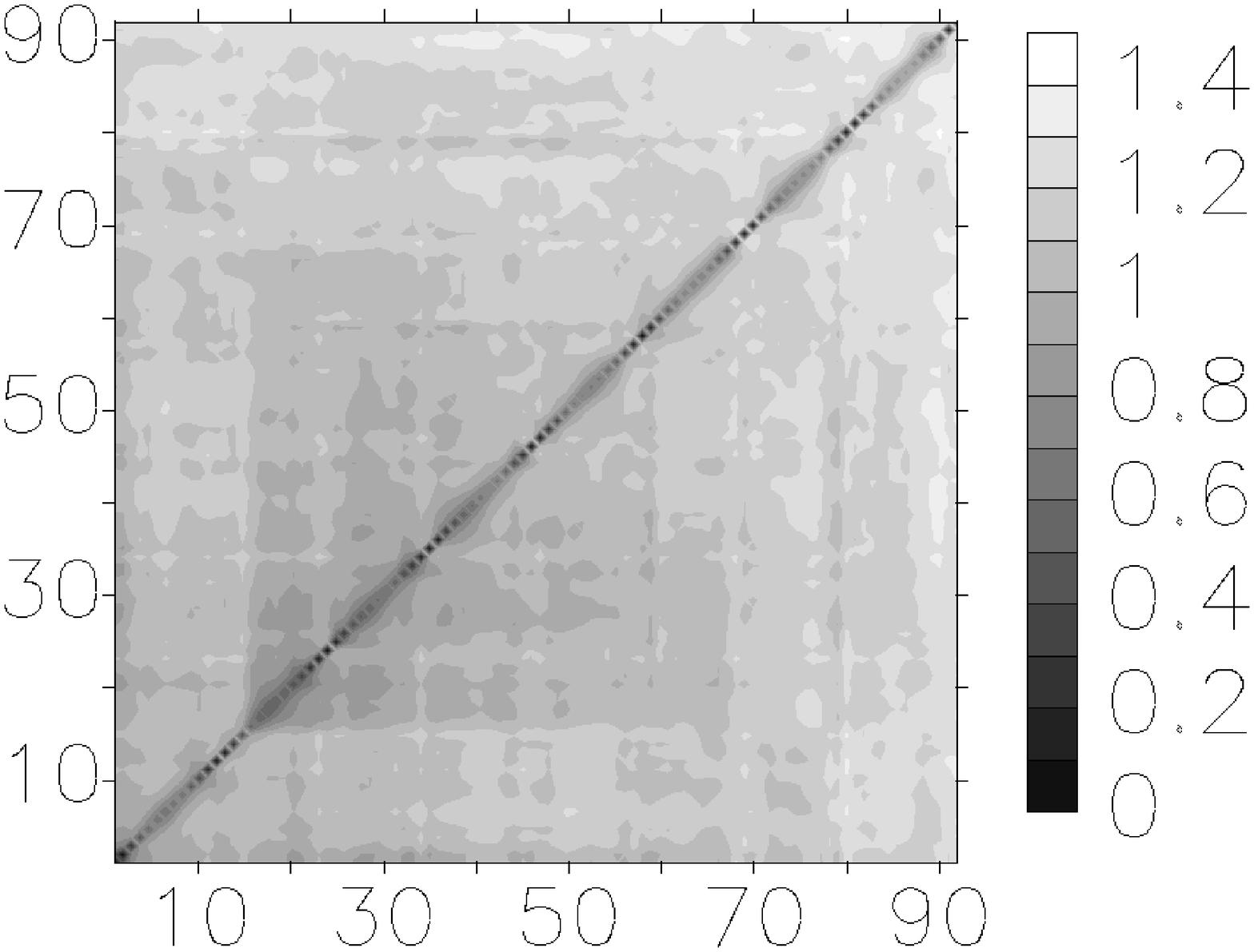}
              \hspace{.05 cm}
              \includegraphics[scale=0.25]{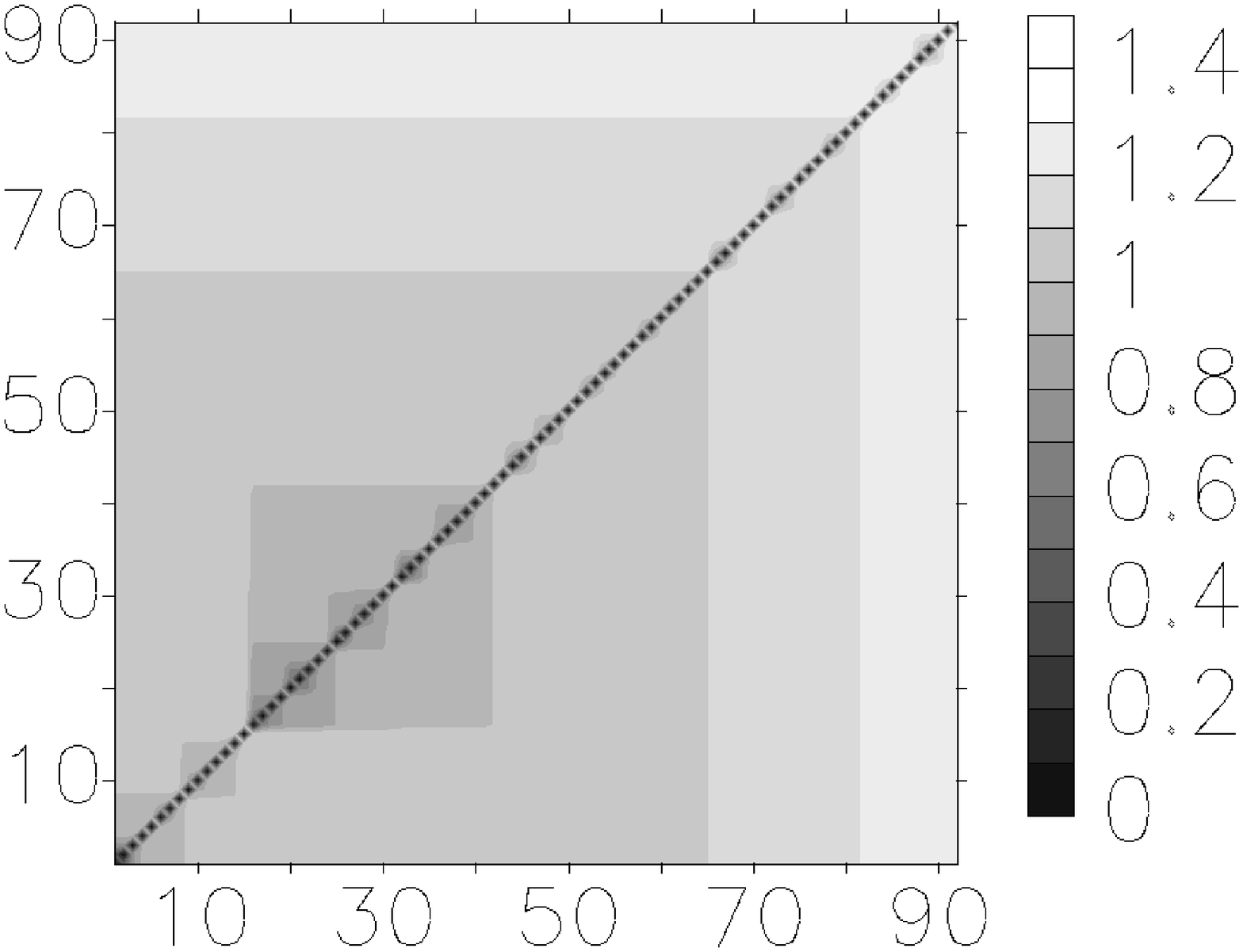}
              }
              \caption{Contour plots of the original correlation matrix (panel a)) and of the one associated to the ultrametric distance (panel b)) obtained by using the ALCA for the daily price returns of 92 highly traded stocks belonging to the SET1 segment of the LSE. Only electronic transactions occurred in year 2002 are considered. Here the stocks are ordered according to the hierarchical tree of Fig. \ref{daydendroaverage}. The figure gives a pictorial representation of the amount of information which is filtered out by using the ALCA.} \label{MATRIX_al_day}
\end{center}
\end{figure}

A comparison of the results obtained by using the SLCA and the ALCA shows a substantial agreement between the output of these two methods. However, a refined comparison shows that the ALCA provides a more structured hierarchical tree. In Fig. \ref{MATRIX_sl_day} and Fig. \ref{MATRIX_al_day} we show a graphical representation of the original correlation matrix 
done in terms of a contour plot. In the contour plot the gray scale represents the values of distances among stocks. In the figure we use as stock order the one obtained by SLCA and ALCA respectively. In both cases we also show the associated ultrametric matrices. A direct comparison of the ultrametric matrices confirms that ALCA is more structured than SLCA. Conversely, the SLCA selects elements of the matrix with correlation values greater than the ones selected by ALCA and then less affected by statistical uncertainty. 

%%%%%%%%%%%%%%%%%%%%%%%%%%%%%%%%%%%%%%%%%%%%%%%%%%%%%%%%%%%%%%%%%%%%%%%%%%%%%%%%%%%%%%%%%%%%%%%%%%%%%%%%%%
\subsection{The Planar Maximally Filtered Graph} \label{PMFG1day}

\begin{figure}
              \includegraphics[width=0.98\textwidth]{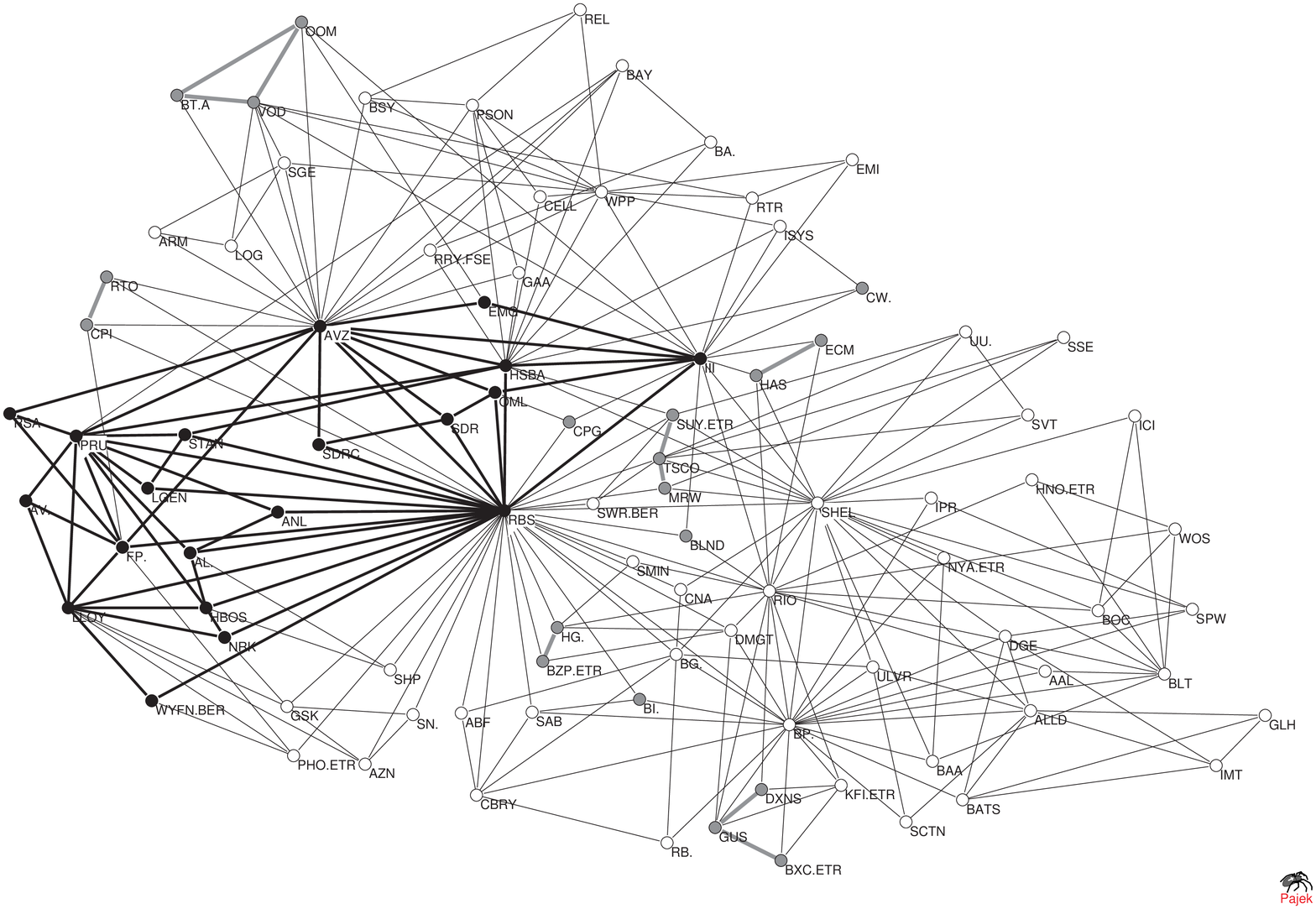}
              \caption{PMFG obtained from daily returns of a set of 92 stocks traded in the LSE in 2002. Black circles are identifying stocks belonging to the Financial sector. Gray circles are identifying stocks belonging to the Services sector. Other stocks are indicated by empty circles. Black thicker lines are connecting stocks belonging to the Financial sector. Gray thicker lines are connecting stocks belonging to the Services sector.}  \label{planarday} 
\end{figure}

In this section we analyze the topological properties of the PMFG of Fig. \ref{planarday} obtained from the distance matrix of daily returns of the stock portfolio. In the figure we again point out the behavior of stocks belonging to the Financial and Services sectors. From the figure we can observe that the Financial sector (black circles) is strongly intra-connected (black thicker edges) whereas for the sector of Services (gray circles) we find just a few intra-sector connections (gray thicker edges). These results agree with the ones observed with the SLCA and the ALCA. The advantage of the study of the PMFG is that, through it, we can perform a quantitative analysis of this behavior. The existence in the graph of completely connected subgraphs, specifically 3-cliques and 4-cliques allows one to investigate the clustering level of sectors through a measure of the intra-cluster connection strength \cite{Tumminello2005}. This measure is obtained by considering a specific sector composed by $n_s$ elements and indicating with $c_4$ and $c_3$ the number of 4-cliques and 3-cliques exclusively composed by elements of the sector. The connection strength $q_s$ of the sector $s$ is therefore defined as

\begin{eqnarray} \label{strength}
          &&    q_s^{4}=\frac{c_4}{n_s-3},\nonumber \\
          &&                       \\      
          &&    q_s^{3}=\frac{c_3}{3 n_s-8}, \nonumber
\end{eqnarray}  
where we distinguish between the connection strength evaluated according to 4-cliques $q_s^{4}$ and 3-cliques $q_s^{3}$. The quantities $n_s-3$ and $3 n_s-8$ are normalizing factors. For large and strongly connected sectors both the measures give almost the same result \cite{Tumminello2005}. When small sectors are considered the quantity $q_s^3$ is more significant than $q_s^4$. Consider for instance a sector of 4 stocks. In this case $q_s^4$ can assume the value 0 or 1, whereas $q_s^3$ can assume one the 5 values 0, 0.25, 0.5, 0.75 and 1, giving a measure of the clustering strength less affected by the quantization error. Note that in the case of $n_s=4$ if $q_s^3$ assumes one of the values 0, 0.25, 0.5 and 0.75 then $q_s^4$ is always 
zero. In Table \ref{tab:PMFG1day} the connection strength is evaluated for all the sectors present in the portfolio. The Financial sector has $q_2^4\cong0.88$ and $q_2^3\cong 0.92$. This last value is second only to the Energy sector (composed by 3 stocks) where all stocks are connected within them so that $q_3^3=1$. The stocks belonging to the Energy sector are BG, BP and SHEL. We see in Fig. \ref{planarday} that both  BP and SHEL are characterized by high values of their degree (number of links with other elements). This fact implies that the Energy sector is strongly connected both within the sector and with other sectors. This behavior is different from what has been observed in the analysis of 100 highly capitalized stocks traded in the US equity market \cite{Tumminello2005}. In Fig. \ref{planarday} we observe that stocks of the Financial sector are strongly connected with stocks belonging to different sectors. In particular RBS is the center of the biggest star in the graph. On the contrary, the sector of Services is poorly intra-connected: $q_8^{4}=0$ and $q_8^{3}\cong 0.02$ and poorly connected to other sectors. In conclusion we observe two different behaviors. The Financial and Energy sectors are strongly intra-connected and strongly connected with other sectors. The sector of Services is poorly intra-connected and poorly interacting with other sectors. 
   
%\begin{widetext}
%\end{widetext}

\begin{table}\nonumber
\begin{center}
\caption{Intra-sector connection strength (daily returns)} \label{tab:PMFG1day}
\vspace{.5 cm}
\begin{tabular}{||l|c|c|c||}
%\tableline
\hline
SECTOR & $n_s$ & $q_s^{4}=c_4/[n_s-3]$ & $q_s^{3}=c_3/[3\,n_s-8]$\\ \hline
%\tableline
%\tableline
Technology            & 4 & $0/1=0$ & $1/4=0.25$ \\
Financial             & 20 & $15/17\cong0.88$ & $48/52\cong0.92$ \\
Energy                & 3 & $-$ & $1/1=1$ \\
Consumer non-Cyclical & 12 & $2/9\cong0.22$ & $8/28\cong0.29$ \\
Consumer Cyclical     & 10 & $1/7\cong0.14$ & $5/22\cong0.23$ \\
Healthcare            &6 & $0/3=0$ & $1/10=0.1$ \\
Basic Materials       & 5 & $0/2=0$ & $3/7\cong0.43$ \\
Services              & 19 & $0/16=0$ & $1/49\cong0.02$ \\
Utilities             & 6 & $0/3=0$ & $0/10=0$ \\
Capital Goods         & 5 & $0/2=0$ & $0/7=0$ \\
Transportation        & 2 & $-$ & $-$ \\ \hline
%\tableline
%\tableline
\end{tabular}
\end{center}
\end{table}

%%%%%%%%%%%%%%%%%%%%%%%%%%%%%%%%%%%%%%%%%%%%%%%%%%%%%%%%%%%%%%%%%%%%%%%%%%%%%%%%%%%%%%%%%%%%%%%%%%%%%%%%%%
%%%%%%%%%%%%%%%%%%%%%%%%%%%%%%%%%%%%%%%%%%%%%%%%%%%%%%%%%%%%%%%%%%%%%%%%%%%%%%%%%%%%%%%%%%%%%%%%%%%%%%%%%%
\section{Empirical Results: 5-minute data} \label{results5}

%%%%%%%%%%%%%%%%%%%%%%%%%%%%%%%%%%%%%%%%%%%%%%%%%%%%%%%%%%%%%%%%%%%%%%%%%%%%%%%%%%%%%%%%%%%%%%%%%%%%%%%%%%
\subsection{Random Matrix Theory} \label{RMT5min}

The properties of correlation matrix and of its eigenvalues and eigenvectors change dramatically when one considers cross correlations between returns computed at a 5-minute time horizon. The largest eigenvalue is $\lambda_1=11.2$ and this sets the variance of the space orthogonal to it to $\sigma^2=0.87$. The noisy region of the spectrum is characterized by the values $\tilde\lambda_{min}=0.78$ and $\tilde\lambda_{max}=0.99$. With these values one would conclude that $19$ eigenvalues contain economic information. This is quite surprising because one would expect that for a short time horizon the correlation coefficients are less influenced by economic sectors than when one considers daily returns. We will see in the following sections that clustering methods support this view. 

\begin{figure}[ptb]
\begin{center}
\includegraphics[scale=.4]{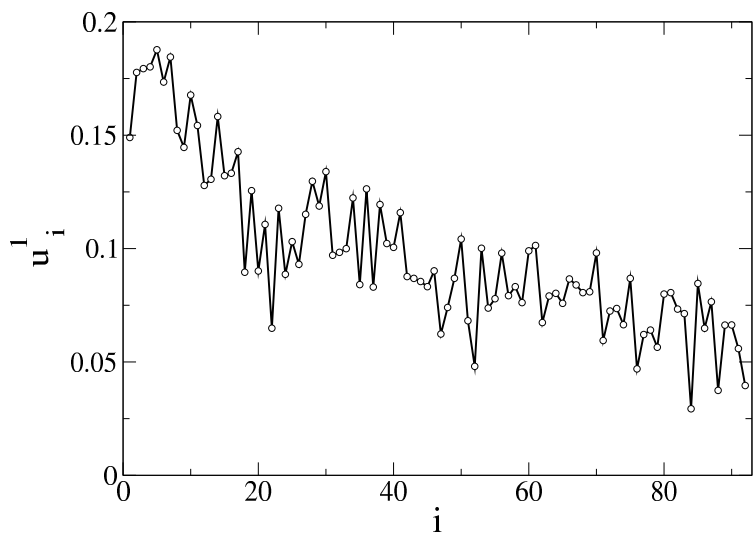}
\caption{
Components ${u}^1_i$ of the first eigenvector of the correlation matrix of 5-minute returns. In the x axis of this figure the stocks are sorted in decreasing order according to the total number of trades recorded in the investigated period.}
\label{1stevec5min}
\end{center}
\end{figure}

Figure~\ref{1stevec5min} shows the components ${u}^1_i$ of the first eigenvector. In the x axis of this figure the stocks are sorted in decreasing order according to the total number of trades recorded in the investigated period.
%, see Fig. \ref{transactions} (circles). 
The figure shows that the most heavily traded stocks have a larger component in the first eigenvector. This behavior is not observed in the first eigenvector for daily returns. 

A possible interpretation of this result is the following. Suppose that, as a first approximation, the dynamics of the set of stocks is described by a one factor model, i.e. a model in which the dynamics of each variable is controlled by a single factor. The equation describing the one factor model is given by
\begin{equation}
                r_i(t)=\gamma_i f(t)+\gamma_i^{(0)}\epsilon_i(t),    \label{onefact}
\end{equation}
where  $\epsilon_i(t)$ is a Gaussian zero mean noise term with unit variance and it is assumed that the noise terms are uncorrelated one with each other and with the factor,
i.e. $\langle \epsilon_i(t)\epsilon_j(t)\rangle=\delta_{ij}$ and $\langle f(t)\epsilon_j(t)\rangle=0$
The parameter $\gamma_i^2$ gives the fraction of variance explained by the common factor $f(t)$ and $\gamma_i^{(0)}=\sqrt{1-\gamma_i^2}$.
The model describes a system where $n$ variables are essentially controlled by a common factor describing a weighted mean. This type of model is, for example consistent with the Capital Asset Pricing Model of stock market behavior. It is possible to show \cite{Lillo2005} that the spectrum of this model is given by a large eigenvalue $\lambda_1\simeq\sum_{i=1}^n\gamma_i^2$ and $n-1$ eigenvalues whose density can be obtained by using RMT. We wish to address here the question of the dependence of the first eigenvector from the $\gamma_i$ parameters. It is possible to show that in the large $n$ limit the first eigenvector is well approximated by the vector ${\bf u}^1\propto {\bf g}\equiv(\gamma_1,\gamma_2,...,\gamma_n)^T$. In fact the correlation matrix of the model of Eq.~\ref{onefact} has off diagonal elements $\rho_{ij}=\gamma_i\gamma_j$ for $i\ne j$ \cite{Lillo2005}. The product of the $i$-th row of the correlation matrix times the vector ${\bf g}$ gives $\gamma_i(1+\sum_{i\ne j} \gamma_j^2)\simeq \gamma_i \sum_{j=1}^n\gamma_j^2 \simeq \gamma_i \lambda_1$, which implies that ${\bf g}$ well approximates the eigenvector ${\bf u}^1$ in the large $n$ limit.

Thus the result shown in Fig.~\ref{1stevec5min} can be interpreted in the following way. At 5-minute horizon the market is approximated by a one factor model of Eq.~\ref{onefact}. The $\gamma_i$ are related to the trading frequencies because more actively traded stocks are usually the ones with the highest capitalization and these stocks are the ones following more closely the mean behavior of the market, i.e. the common factor $f(t)$.

\begin{figure}[ptb]
\begin{center}
\includegraphics[scale=.3]{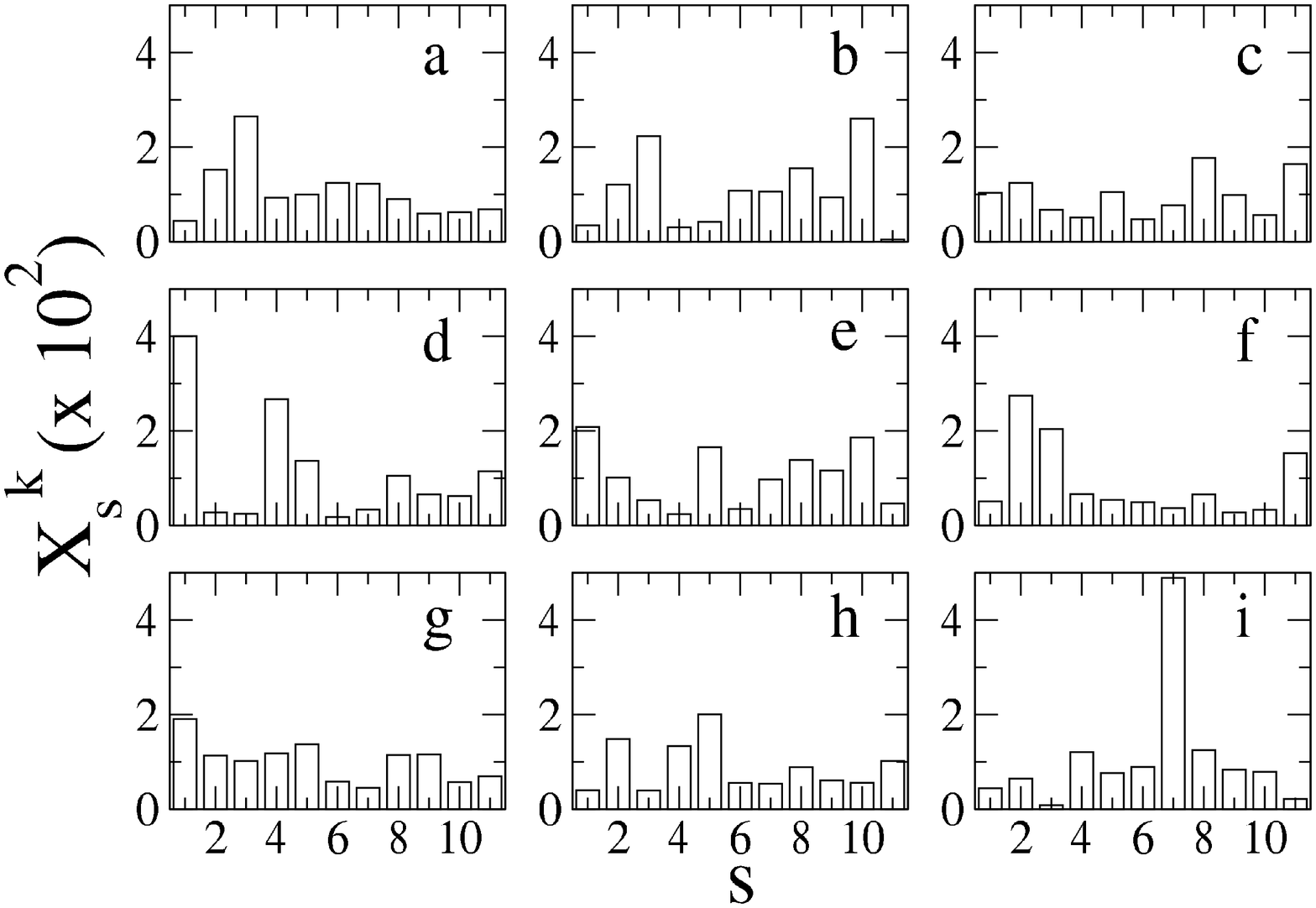}
\caption{
Contribution  $X^k_s$ of Eq.~\ref{projectioneq} for the first $9$ eigenvectors of the correlation matrix of 5-minute returns of $92$ LSE stocks. The order of sectors is the same as in Table \ref{classification}.}
\label{projection5min}
\end{center}
\end{figure}

The sector analysis of 5-minute correlation matrix performed with RMT shows less clear results than for daily returns. Figure~\ref{projection5min} shows the contribution  $X^k_s$ of Eq.~\ref{projectioneq} for the first $9$ eigenvalues of the correlation matrix of 5-minute returns of $92$ LSE stocks. The first, second, fourth, sixth, and especially ninth eigenvector show peaks indicating the prominent role of one or few sectors in determining the dynamics of these eigenvectors. 

However, a systematic correspondence as in the case of daily returns is not observed. Moreover it is unclear what kind of information can be associated to the first 19 eigenvalues carrying information not affected by statistical uncertainty. 

It is therefore worth to consider what results are provided by correlation based clustering algorithms for the same time horizon.

%%%%%%%%%%%%%%%%%%%%%%%%%%%%%%%%%%%%%%%%%%%%%%%%%%%%%%%%%%%%%%%%%%%%%%%%%%%%%%%%%%%%%%%%%%%%%%%%%%%%%%%%%%
\subsection{Single Linkage Correlation based Clustering} \label{SLCA5min}

\begin{figure}
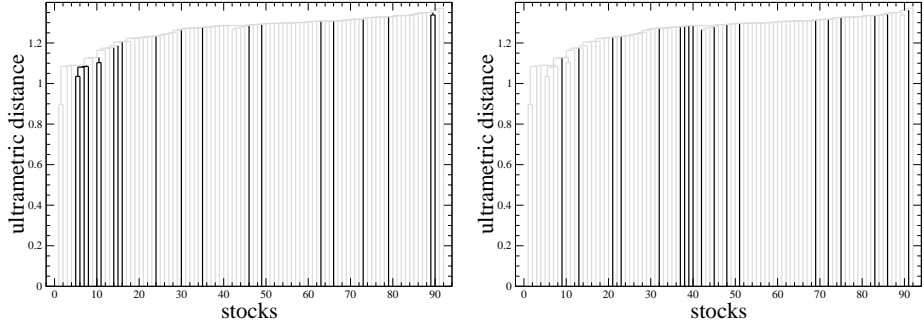

\begin{center}
              \hbox{
              \includegraphics[scale=0.25]{HT_sl_5min_financial.eps}
              \hspace{.05 cm}
              \includegraphics[scale=0.25]{HT_sl_5min_services.eps}
              }
\end{center}
              \caption{hierarchical tree obtained by using the SLCA starting from the 5-minute price returns of 92 highly traded stocks belonging to the SET1 segment of the LSE. Only electronic transactions occurred in year 2002 are considered. In panel a) the Financial economic sector is highlighted. In panel b) the Services economic sector is highlighted.} \label{HT_sl_5min}
\end{figure}

At a 5-minute time horizon the structure of the MST and hierarchical tree are quite different from the analogous trees at a daily time horizon. Figure \ref{HT_sl_5min} shows the hierarchical tree obtained by using the SLCA for the selected 92 stocks at a 5-minute time horizon. We proceed here in analogy with the discussion done for the one day time horizon to put in emphasis similarities and differences between the results obtained for the two time horizons. Specifically, in panel a) all stocks belonging to the Financial economic sector are highlighted, while in panel b) the stocks belonging to the Services economic sector are highlighted. The hierarchical tree shows that now the mean level of correlation in the market is lower than at one day time horizon. The level of clustering is also less pronounced at this time horizon. In fact, panel a) of Fig. \ref{HT_sl_5min} shows how the stocks of the Financial sector are only poorly clustered, contrary to the case shown in panel a) of Fig. \ref{HT_sl_day}. Panel b) of Fig. \ref{HT_sl_5min} shows that at a 5-minute time horizon there is absence of any amount of clustering for stocks of the Services sector. 

\begin{figure}
\begin{center}
              \includegraphics[scale=0.35]{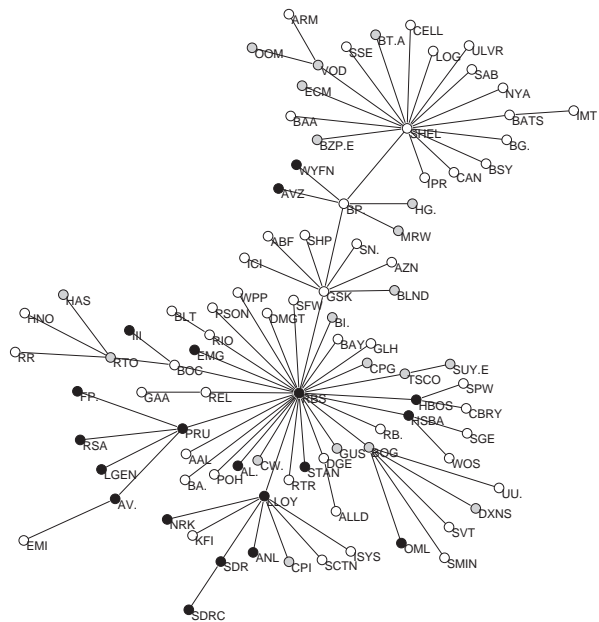}
              \caption{MST obtained starting from the 5-minute price returns of 92 highly traded stocks belonging to the SET1 segment of the LSE. Only electronic transactions occurred in year 2002 are considered. The Financial economic sector (black) and the Services (gray) economic sector are highlighted. It is evident the existence of two stocks that behave as hubs. One of them is RBS, which gathers 14 stocks, 10 of which belong to the Financial sector. The other hub is SHEL which gathers 10 stocks.} \label{MST_sl_5min}
\end{center}
\end{figure}

In Fig. \ref{MST_sl_5min} the MST of the 92 stocks computed at a 5-minute time horizon is shown. As in Fig. \ref{MST_sl_day}, the stocks belonging to the Financial sector (black circles) and Services sector (gray circles) are highlighted. Several stocks of the Financial sector cluster around RBS. The organization of the 92 stocks around two hubs (SHEL and RBS) is here more pronounced than at a daily time horizon. In particular, RBS has now a degree of 29 and SHEL has a degree of 17. However, while at a daily time horizon 10 stocks of the Financial sector are linked to RBS, at the present time horizon only 7 Financial stocks are linked to RBS. A possible interpretation is that RBS acts as hub mainly for its economic sector at a daily time horizon, while at a shorter time horizon, when economic sectors are expected to play a minor role, RBS is influential for the whole stock market. These results are similar to what has been observed for 100 stocks traded in US equity markets in Ref. \cite{Bonanno2001}.

%%%%%%%%%%%%%%%%%%%%%%%%%%%%%%%%%%%%%%%%%%%%%%%%%%%%%%%%%%%%%%%%%%%%%%%%%%%%%%%%%%%%%%%%%%%%%%%%%%%%%%%%%%
\subsection{Average Linkage Correlation based Clustering} \label{ALCA5min}

\begin{figure}
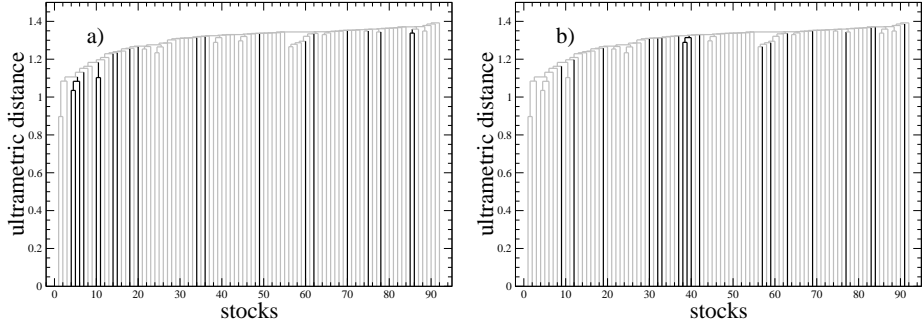
 
\begin{center} 
              \hbox{ 
              \includegraphics[scale=0.25]{HTDistAL_LSE92_2002_5_financial} 
              \hspace{.05 cm} 
              \includegraphics[scale=0.25]{HTDistAL_LSE92_2002_5_services} 
              } 
              \caption{Dendrogram associated to the ALCA performed on 5-minute returns of a portfolio of 92 stocks traded in the LSE in 2002. Panel a): The black lines are identifying stocks belonging to the Financial Sector. Panel b): The black lines are identifying stocks belonging to the Services Sector} 
\label{5mindendroaverage} 
\end{center} 
\end{figure}

In Fig. \ref{5mindendroaverage} we show the dendrogram obtained for the 5-minute returns by applying the ALCA to the correlation based distance matrix of the system. In panel a) of Fig. \ref{5mindendroaverage} the black lines are again identifying the Financial sector. In the figure, we observe that just an intra-sector cluster of 3 elements is formed. Specifically, {\em{Lloyds TSB Group}} (LLOY), RBS, {\em{HSBC Holdings}} (HSBA) cluster together at a distance level $d \sim 1.08$. In panel b) the black lines are identifying stocks belonging to the Services sector. As in the case of daily returns only an intra-sector cluster of 3 stocks is recognized by the ALCA. It involves stocks {\em{Dixons Group}} (DXNS), {\em{Boots}} (BOG) and {\em{Compass Group}} (CPG).

A direct comparison of Fig. \ref{5mindendroaverage} and Fig. \ref{daydendroaverage} shows that at the time horizon of 5-minute the Financial cluster observed for daily returns is not yet formed. More generally a strong reduction of structures in the dendrogram is observed when going from daily returns to 5-minute returns. 

%%%%%%%%%%%%%%%%%%%%%%%%%%%%%%%%%%%%%%%%%%%%%%%%%%%%%%%%%%%%%%%%%%%%%%%%%%%%%%%%%%%%%%%%%%%%%%%%%%%%%%%%%%
\subsection{The Planar Maximally Filtered Graph} \label{PMFG5min}

\begin{figure}
              \includegraphics[width=0.98\textwidth]{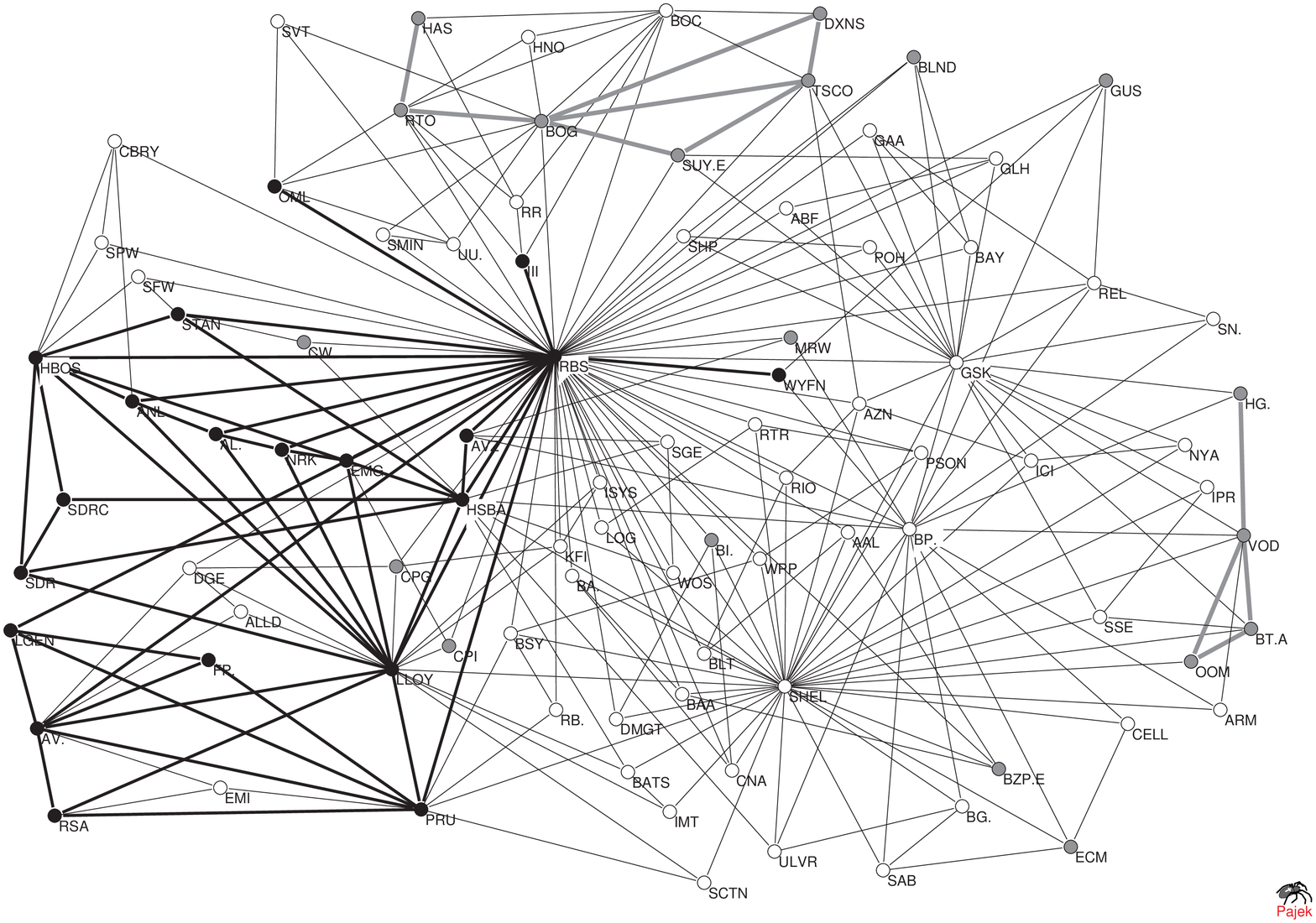}
              \caption{PMFG obtained from 5-minute returns of a set of 92 stocks traded in the LSE in 2002. Black circles are identifying stocks belonging to the Financial sector. Gray circles are identifying stocks belonging to the Services sector. Other stocks are indicated by empty circles. Black thicker lines are connecting stocks belonging to the Financial sector. Gray thicker lines are connecting stocks belonging to the Services sector.}  \label{planar5mins} 
\end{figure}

Lastly we discuss the properties of the PMFG obtained for the portfolio of stocks by considering 5-minute returns. A comparison of Fig. \ref{planar5mins} and Fig. \ref{planarday} shows that the PMFG experiences a major modification. In fact, if we just focus on the stocks with the highest value of degree, some of them increase their degree whereas others decrease their own. Specifically, RBS and SHEL increase their degree from 42 to 62 and from 24 to 37 respectively, whereas BP and {\em{Amvescap}} (AVZ) decrease their degree from 23 to 18 and from 24 to 5 respectively. This difference shows that the role of most connected stock can be quite different at different time horizons.

In Table \ref{tab:PMFG5min} the intra-sector connection strength discussed in section \ref{PMFG1day} is evaluated for the 5-minute time horizon. Table \ref{tab:PMFG5min} shows that only three sectors have a connection strength different from zero. Specifically the Energy sector has connection strength $q_3^3=1$ and the Financial sector $q_2^4\cong0.71$ and $q_2^3\cong0.75$ indicating a behavior of both the sectors similar to the one observed for daily returns. Finally the sector of Services has $q_8^{3}\cong0.06$ revealing a clustering of the same order of the one observed for daily returns. A critical difference between the two time horizons appears for several sectors. The most striking example being the sector of Basic Materials. In Table \ref{tab:PMFG5min} we see that the connection strength of the sector is zero, with respect to both the connection strength measures. On the contrary when daily returns are considered the connection strength $q_7^3=0.43$ was observed. This difference suggests that the intra-sector correlation of Basic Materials stocks needs time to be settled up into the market. Several of the remaining sectors show a behavior analogous to the one of Basic Materials. This effect is detected by all the considered techniques, thus indicating the need of time for the market to assess a certain degree of correlation among stocks.

\begin{table}\nonumber
\begin{center}
\caption{Intra-sector connection strength (5-minute returns)} \label{tab:PMFG5min}
\vspace{.5 cm}
\begin{tabular}{||l|c|c|c||}
%\tableline
\hline
SECTOR & $n_s$ & $q_s^{4}=c_4/[n_s-3]$ & $q_s^{3}=c_3/[3\,n_s-8]$\\\hline
%\tableline
%\tableline
Technology            & 4 & $0/1=0$ & $0/4=0$ \\
Financial             & 20 & $12/17\cong0.71$ & $39/52\cong0.75$ \\
Energy                & 3 & $-$ & $1/1=1$ \\
Consumer non-Cyclical & 12 & $0/9=0$ & $0/28=0$ \\
Consumer Cyclical     & 10 & $0/7=0$ & $0/22=0$ \\
Healthcare            & 6 & $0/3=0$ & $0/10=0$ \\
Basic Materials       & 5 & $0/2=0$ & $0/7=0$ \\
Services              & 19 & $0/16=0$ & $3/49\cong0.06$ \\
Utilities             & 6 & $0/3=0$ & $0/10=0$ \\
Capital Goods         & 5 & $0/2=0$ & $0/7=0$ \\
Transportation        & 2 & $-$ & $-$ \\ \hline
%\tableline
%\tableline
\end{tabular}
\end{center}
\end{table}

%%%%%%%%%%%%%%%%%%%%%%%%%%%%%%%%%%%%%%%%%%%%%%%%%%%%%%%%%%%%%%%%%%%%%%%%%%%%%%%%%%%%%%%%%%%%%%%%%%%%%%%%%%
%%%%%%%%%%%%%%%%%%%%%%%%%%%%%%%%%%%%%%%%%%%%%%%%%%%%%%%%%%%%%%%%%%%%%%%%%%%%%%%%%%%%%%%%%%%%%%%%%%%%%%%%%%
\section{Conclusions}

All the methods considered in the present paper are able to detect information about economic sectors of stocks starting from the synchronous correlation coefficient matrix of return time series. The degree of efficiency in the detection is depending on the return time horizon. Specifically, the system is more hierarchically structured at daily time horizons confirming that the market needs a finite amount of time to assess the correct degree of cross correlation between pairs of stocks whose prices are simultaneously recorded \cite{Bonanno2001}. Our comparative study shows that, at a given time horizon, the considered methods can provide different information about the system. For example, at one day time horizon the method based on RMT predominantly associates the eigenvectors of the six highest eigenvalues which are not affected by statistical uncertainty respectively to  the market factor (first eigenvalue and eigenvector), the Consumer non-Cyclical sector (second eigenvalue), the Financial sector (third eigenvalue), a linear combination of Technology and Capital Goods sectors (fourth and fifth eigenvalues) and the Helthcare sector (sixth eigenvalue). In the present case, RMT does not provide information about the existence and strength of economic relation between stocks belonging to the sectors of Energy, Consumer Cyclical, Basic Materials, Services, Utilities and Transportation. A detailed investigation of the hierarchical trees obtained by the SLCA and ALCA shows that these methods are able to detect efficiently most of the clusters detected with the methods of the RMT and also other clusters related to other sectors. The only sector that RMT detects in a way which is more efficient with respect to the correlation based clustering procedures is the cluster of stocks belonging to Consumer non-Cyclical sector. One sector which is not detected by both RMT and hierarchical clustering methods is the sector of Services. RMT is not able to detect it whereas SLCA and ALCA are able to detect only limited aggregation of elements of it. 

Our comparative analysis of the hierarchical clustering methods shows that SLCA and ALCA also provide different information. Specifically, the SLCA is providing information about the highest level of correlation of the correlation matrix whereas the ALCA averages this information within each considered cluster. In this way the average linkage clustering is able to provide a more structured information about the hierarchical organization of the stocks of a portfolio.

Additional information with respect to the one associated with the MST of the system can be also detected by considering the properties of the PMFG. This graph provides quantitative information about the degree of inter-cluster and intra-cluster connection of the various elements.  

In summary, we believe that our empirical comparison of different methods provide an evidence that RMT and hierarchical clustering methods are able to point out information present in the correlation matrix of the investigated system. The information that is detected with these methods is in part overlapping but in part specific to the selected investigating method. In short, all the approaches detect information but not exactly the same one. For this reason an approach that simultaneously makes use of several of these methods may provide a better characterization of the investigated system than an approach based on just one of them. 

{\bf Acknowledgments} FL and RNM acknowledge support from ``Cost P10: Physics of Risk''. Authors acknowledge support from the research project MIUR 449/97 ``High frequency dynamics in financial markets", the research project MIUR-FIRB RBNE01CW3M ``Cellular Self-Organizing nets and chaotic nonlinear dynamics to model and control complex system and from the European Union STREP project n. 012911 ``Human behavior through dynamics of complex social networks: an interdisciplinary approach".

%%%%%%%%%%%%%%%%%%%%%%%%%%%%%%%%%%%%%%%%%%%%%%%%%%%%%%%%%%%%%%%%%%%%%%%%%%%%%%%%%%%%%%%%%%%%%%%%%%%%%%%%%%
%%%%%%%%%%%%%%%%%%%%%%%%%%%%%%%%%%%%%%%%%%%%%%%%%%%%%%%%%%%%%%%%%%%%%%%%%%%%%%%%%%%%%%%%%%%%%%%%%%%%%%%%%%

\end{document}